\def\Title{MEASURING PROCESSES AND REPEATABILITY HYPOTHESIS }
\def\Author{Masanao Ozawa}
\documentclass[12pt]{article}

\usepackage{times}
\evensidemargin=\oddsidemargin
  \let\ps=\psi
  \newcommand{\beq}{\begin{equation}}
  \newcommand{\eeq}{\end{equation}}
  \newcommand{\beql}[1]{\begin{equation}\label{eq:#1}}

  \newcommand{\beas}{\begin{eqnarray*}}
  \newcommand{\eeas}{\end{eqnarray*}}
  \newcommand{\beqa}{\begin{eqnarray}}
  \newcommand{\eeqa}{\end{eqnarray}}
  \newcommand{\beqas}{\begin{eqnarray*}}
  \newcommand{\eeqas}{\end{eqnarray*}}
  \newtheorem{Theorem}{Theorem}[section]
  
  \newtheorem{Lemma}[Theorem]{Lemma}

  \newenvironment{Proof}{\begin{trivlist}
    \item[\hskip \labelsep {\em \indent Proof.}]}{\qed\end{trivlist}}
 \newcommand{\qed}{{\em QED}}
  \newcommand{\R}{{\bf R}}
 \newcommand{\cB}{{\cal B}}
  \newcommand{\cE}{{\cal E}}
  \newcommand{\cH}{{\cal H}}
  \newcommand{\cK}{{\cal K}}
  \newcommand{\cL}{{\cal L}}
  \newcommand{\cW}{{\cal W}}
  \newcommand{\al}{\alpha}
  \newcommand{\be}{\beta}
  
  \newcommand{\de}{\delta}
  
  \newcommand{\et}{\eta}

  \newcommand{\la}{\lambda}
  \newcommand{\nn}{\nonumber}
  
  \newcommand{\ph}{\phi}
  \newcommand{\rh}{\rho}
  \newcommand{\si}{\sigma}

  \newcommand{\Ps}{\Psi}
  \newcommand{\Tr}{\mbox{\rm Tr}}
\newcommand{\bracket}[1]{\langle#1\rangle}
  \newcommand{\ketbra}[1]{|#1\rangle\langle #1|}
\newcommand{\bTheorem}{\begin{Theorem}}
\newcommand{\eTheorem}{\end{Theorem}}
\newcommand{\bLemma}{\begin{Lemma}}
\newcommand{\eLemma}{\end{Lemma}}
\newcommand{\bProof}{\begin{Proof}}
\newcommand{\eProof}{\end{Proof}}
\newcommand{\oR}{\overline{\R}}

  \title{{\bf \Title}}
  \author{\sc\Author \\
  \small\em Department of Mathematics, College of General Education\\
  \small\em  Nagoya University, Chikusa-ku, Nagoya 464-8601, Japan}
  \date{}
 
\begin{document}
\maketitle
\begin{abstract}
Srinivas [Commun.\ Math.\ Phys.\ 71 (1980), 131--158] proposed
a postulate in quantum mechanics that extends the von Neumann-L\"{u}ders 
collapse postulate to observables with continuous spectrum.
His collapse postulate does not determine a unique state change, 
but depends on a particular choice of an invariant mean.  
To clear the physical significance of employing different invariant means, 
we construct different measuring processes of the same observable satisfying 
the Srinivas collapse postulate corresponding to any given invariant means.  
Our construction extends the von Neumann type measuring process 
with the meter being the position observable 
to the one with the apparatus prepared in a non-normal state.
It is shown that the given invariant mean corresponds to the momentum 
distribution of the apparatus in the initial state, which is
determined as a non-normal state, called a Dirac state, such that 
the momentum distribution is the given invariant mean and that
the position distribution is the Dirac measure. 
\end{abstract}

\section{Introduction}
\label{se:1}
\setcounter{equation}{0}
The problem of extending the von Neumann-L\"{u}ders collapse postulate [4,5] to observables with continuous spectrum is one of the major problems of the quantum theory of measurement. Recently. Srinivas [11] posed a set of postulates which gave an answer to this problem. However. it does not seem to be a complete solution. The following two problems remain. 

(1) 
The Srinivas collapse postulate is not consistent with the $\sigma$-additivity of probability distributions and it requires ad hoc treatment of calculus of probability 
and expectation. How can we improve his set of postulates in order to retain the consistency with the $\sigma$-additivity of probability?

(2) 
His collapse postulate depends on a particular choice of an invariant mean. What is the physical significance of employing different invariant means? Can we characterize the various different ways of measuring the same observable [11;p.149]?

The purpose of this paper is to resolve the second question by constructing different measuring processes of the same observable satisfying the Srinivas collapse postulate corresponding to the given invariant means. In our construction, the pointer position of the apparatus is the position observable and the given invariant mean corresponds to the momentum distribution at the initial state of the apparatus. Thus the choice of the invariant mean characterizes the state preparation of the apparatus. 

For the general theory of quantum measurements of continuous observables, we shall refer to Davies [1], Holevo [3] and Ozawa [6--10]. The entire discussion including the solution of the first question above will be published elsewhere. 

\section{Formulation of the problem}

 In this paper, we shall deal with quantum systems with finite degrees of freedom. In the conventional formulation, the states of a system are represented by density operators on a separable Hilbert space $\cH$ and the observables are represented by self-adjoint operators on $\cH$. In this formulation, however, as shown in [7;Theorem 6.6], we cannot construct measuring processes satisfying the repeatability hypothesis, which follows from the Srinivas postulates; hence some generalization of the framework of quantum mechanics is necessary. We adopt the formulation that the states of a system are represented by norm one positive linear functionals on the algebra $\cL (\cH)$ of bounded operators on $\cH$; states 
corresponding to density operators will be called {\em normal states}. 
For any state $\sigma$ and compatible observables $X, Y$ we shall denote by 
$\Pr\{X\in dx, Y\in dy\|\si\}$ the joint distribution of the outcomes of the simultaneous measurement of $X$ and $Y$. Our basic assumption is that 
$\Pr\{X\in dx,Y\in dy\|\si\}$ is a $\si$-additive probability distribution on $\oR^{2}$ uniquely determined by the relation 
\beql{(2.1)}
\int_{\oR^{2}} f(x,y) \Pr\{X\in dx,Y\in dy\|\si\} = \bracket{f(X,Y),\si}, 
\eeq
for all $f\in C(\oR^2)$, where $\oR=\R\cup\{+\infty\}\cup\{-\infty\}$
and $C(\oR^2)$  stands for the space of continuous functions on $\oR^2$. 
If $\si$ is a normal state, Eq.(2.1) is reduced to the usual statistical formula.
Apart from classical probability theory, we can consider another type 
of joint distributions in quantum mechanics. 
Let $\bracket{X, Y}$ be an ordered pair of any observables. We shall denote by $\Pr\{X\in dx; Y\in dy\|\si\}$ the joint distribution of the outcomes of the successive measurement of $X$ and $Y$, performed in this order, in the initial state $\si$. Let $\eta$ be a fixed invariant mean on the space $CB(R)$ of continuous bounded functions on $\R$. 
Let $X$ be an observable. Denote by $\cE^X_\eta$ the norm one projection from $\cL(\cH)$ onto $\{X(B); B\in \cB(\R)\}'$ such that 
\beql{(2.2)}
\Tr[\cE^{X}_{\eta}[A]\rho]= \et_u\Tr[e^{iuX}Ae^{-iuX}\rho], 
\eeq
for all normal state $\rho$ and $A\in \cL(\cH)$, where $\cB(\R)$ stands for the Borel $\si$-field of $\R$,
$X(B)$ stands for the spectral projection of $X$ for $B\in \cB(\R)$,
and $'$ stands for the operation making the commutant in $\cL(\cH)$. Then by a slight modification, the Srinivas collapse postulate asserts the following relation for the successive measurement of $X$ and any bounded observable $Y$:
\beql{(2.3)}
\int_{\R}y \Pr\{X\in B; Y\in dy\|\rho\} = \Tr [X (B)\cE^X_{\eta}[Y] \rho],
\eeq
for all normal state $\rho$ and $B\in B(\R)$. Obviously, this relation implies the following generalized Born statistical formula [11]: If $X$ and $Y$
are compatible then 
\beql{(2.4)}
\Pr\{X\in B;Y\in C\|\rho\} = \Tr\{X(B)Y(C)\rho\},  
\eeq
for all normal state $\rho$ and $B, C\in \cB(\R)$. Our purpose is to construct a measuring process of $X$ which satisfies the Srinivas collapse postulate 
Eq.(2.3). 

Throughout this paper, we shall fix an invariant mean $\et$ which is, by a technical reason, a topological invariant mean on $CB(\R)$ (cf. [2; p.24]).

\section{Dirac state}

In this section, we shall consider a quantum system with a single degree of freedom. Denote by $Q$ the position observable and by $P$ the momentum observable. 
A state $\de$ on $\cL(L^2(\R))$ is called an {\em $\et$-Dirac state} 
if it satisfies the following conditions (D1)--(D2):
\begin{itemize}
\item[(D1)] For each $f \in  CB(\R)$, ${\langle}f(Q),\de{\rangle} = f(0)$.
\item[(D2)] For each $f \in  CB(\R)$, ${\langle}f(P),\de{\rangle} = \et(f)$.
\end{itemize}

\bLemma
For any $f\in CB(\R)$, $\cE^Q_\et(f (P) ) = \et(f)1$. 
\eLemma
\bProof  Let $\xi$ be  a  unit  vector  in  $L^2(\R)$.  Let
$g(p) = |\xi(-p)|^2$. Then $g$  is  a  density  function  on  $\R$.  
For  any  $f\in CB(\R)$,  
we  have  
\beqas
\bracket{\xi|\cE^Q_\et(f(P))|\xi}
&=&\et_u \bracket{\xi|e^{iuQ}f(P) e^{-iuQ}|\xi}\\ 
&=&\et_u \bracket{\xi|f(P + u1)|\xi}\\
&=&\et_u \int_{\R} f (p + u)|\xi(p)|^2 dp\\ 
&=&\et_u (f *g)(u) = \et(f),  
\eeqas
where $f *g$ stands for the convolution of $f$ and $g$. 
It follows  that $\cE^{Q}_\et(f(P)) =\et(f)1$.
\eProof
 
\bTheorem
For every topologically invariant mean $\et$,
there exists an $\et$-Dirac state.
\eTheorem
\begin{Proof}  
Let $\ph$ be a state on $\{Q(B) ; B\in\cB(\R)\}'$ such that 
${\langle}f(Q),\ph{\rangle} =f(0)$ for all $f\in CB(\R)$
and $\de$ a state on $\cL(L^2(\R))$
such that  ${\langle}A,\de{\rangle} = {\langle}\cE^{Q}_{\et}(A),\ph{\rangle}$ for all $A \in  \cL(L^2(\R))$.  
Then by
Lemma 3.1, $\de$ is obviously an $\et$-Dirac state.
\end{Proof}

\section{Canonical measuring processes}
\setcounter{equation}{0}

Let $X$ be an observable of a quantum system I described by a Hilbert space $\cH$. 
We  consider the following measuring process of $X$ by an apparatus system II. The apparatus system II is a system with a single degree of freedom described by the Hilbert space $\cK = L^2(\R)$. 
Thus the composite system I+II is described by the Hilbert space $\cH\otimes \cK$, which will be identified with the Hilbert space 
$L^2(\R;\cH)$ of all norm square integrable $\cH$-valued functions 
on $\R$ by the Schr\"{o}dinger representation of $\cK$. The pointer position of the apparatus system is the position observable $Q$. The interaction between the measured system I and the apparatus system II is given by the following Hamiltonian: 
\beql{(4.I)}
H_{int}=\la(X\otimes P) ,
\eeq 	
where $P$ is the momentum of the apparatus. The strength $\la$ of the interaction is assumed to be sufficiently large that other terms in the Hamiltonian can be ignored. Hence the Schr\"{o}dinger equation 
will be $(h= 2\pi)$ 
\beql{(4.2)}
\frac{\partial}{\partial t}\Psi_t(q)=
-\la\left(X\otimes \frac{\partial}{\partial t}\right)\Psi_t(q), 
\eeq
in the q-representation, where $\Ps_t\in\cH\otimes\cK$. 
The measurement is carried out by the interaction during a finite time interval from $t= 0$ to $t = 1/\la$. The outcome of this measurement is obtained by the measurement of $Q$ at time $t =1/\la$.  
The statistics of this measurement depends on the initially 
prepared state $\si$ of the apparatus. According to [7;Theorem 6.6], 
if $\si$ is a normal state then this measurement cannot satisfy Eq.~(2.3). Now we assume that the initial state of the apparatus is an 
$\et$-Dirac state $\de$ and we shall call this measuring process as a 
{\em canonical measuring process} of $X$ 
with preparation $\de$. 

In order to obtain the solution of Eq.~(4.2), assume the initial 
condition
\beql{(4.3)}
\Ps_0=\ps\otimes \al,
\eeq
where $\ps\in\cH$ and $\al\in\cK$.
The solution of the Schr\"{o}dinger equation 
is given by 
\beql{(4.4)}
\Ps_t=e^{-it\la (X\otimes P)}\ps\otimes \al,
\eeq
and hence for any $\ps\in\cH$ and $\be\in\cK$, we have
\beqas 
\lefteqn{
\int_{\R}{\langle}\ph\otimes \be(q)|\Ps_t(q){\rangle}\, dq}\quad \\
&=&
\int_{\R^2}e^{-it\la x p}
{\langle} \ph\otimes\be|X(dx)\otimes P(dp)|\ps\otimes \al{\rangle}\\
&=&
\int_{\R}
{\langle} \be|e^{-it\la xP}|\al{\rangle}\bracket{\ph|X(dx)|\ps}\\
&=&
\int_{\R}
(\int_{\R} \be(q)^*\al(q-t\la x)dq )\bracket{\ph|X(dx)|\ps}\\
&=&
\int_{\R^2}\al(q1 -t\la x){\langle}\be(q)\ph|X(dx)|\ps{\rangle}\, dq\\
&=&
\int_{\R}
{\langle}\ph\otimes \be(q)|\al(q -t\la X)|\ps{\rangle}\, dq.
\eeqas

It follows that
\beql{(4.5)}
\Ps_t(q)=\al(q1 -t\la X)\ps.
\eeq
For $t =1/\la$, we have
\beql{(4.6)}
\Ps_{1/\la}(q)=\al(q1 -X)\ps.
\eeq

\begin{Theorem} For any $f\in L^{\infty}(\R)$, we have
\beql{(4.7)}
U^{*}_{t}(1\otimes f(Q))U_t=f(t\la (X\otimes 1)+1\otimes Q),
\eeq
where $U_t=e^{-it\la(X\otimes P)}$.
\end{Theorem}
\bProof
By Eq.~(4.5), for any $\ps\in\cH$  and $\al\in\cK$, we have
\beqas
\lefteqn{
\bracket{\ps\otimes\al|U_t^*(1\otimes f(Q))U_t|\ps\otimes\al}}\quad\\
&=&\int_{\R} f(q)  {\langle}\ps|\al(q1-t\la X)^*\al(q1 -t\la X)|\ps{\rangle}\, dq\\
&=&\int_{\R^2} f(q)|\al(q-t\la x)|^2\, dq\bracket{\ps|X(dx)|\ps}\\
&=&\int_{\R^2} f(q+t\la x)|\al(q)|^2\, dq\bracket{\ps|X(dx)|\ps}\\
&=& {\langle}\ps\otimes \al|f(t\la (X\otimes 1)+1\otimes Q)|\ps\otimes \al{\rangle}.
\eeqas
Thus the assertion holds.
\eProof

\section{Statistics of Measurement}
\setcounter{equation}{0}
Suppose that the state of the measured system I at $t =0$ 
is a normal state $\rh$. We shall denote by $\rh\otimes \de$
the state at $t =0$ of the composite system I+II, 
which is defined by the relation $\bracket{T,\rho\otimes \de} = {\langle}\cE_{\rh}(T),\de{\rangle}$ 
for all $T\in  \cL(\cH)\otimes\cL(\cK)$, where 
$\cE_{\rh}: \cL(\cH)\otimes \cL(\cK)\to\cL(\cK) $
is a normal completely positive map such that
$\cE_\rh(A_1\otimes A_2) = \Tr[A_1\rh]A_2$ for all $A_1\in \cL(\cH)$.
Thus letting $U = e^{-i (X\otimes P)}$, the state at $t = 1/\la$
of the composite system I+II is $U(\rh\otimes \de)U^*$.
 
Let $Y$ be a bounded observable of the system I. By the argument similar with [7;\S.3], the joint distribution $\Pr\{X\in dx;Y\in dy\|\rh\}$ of the outcomes of the successive measurement of $X$ and $Y$ 
coincides with the joint distribution 
$\Pr\{Q\in dq, Y\in dy\| U(\rh\otimes\de)U^*\}$ of the simultaneous measurement of the pointer position $Q$ and $Y$ at time $t =1/\la$, 
i.e.,
\beql{(5.1)} 
\Pr\{X\in dx;Y\in dy\|\rh\}= 
\Pr\{Q\in dx, Y\in dy\|U(\rh\otimes \de)U^*\}.
\eeq

The rest of this section will be devoted to proving Eq.~(2.3) for this measuring process. Denote by $\cE_{\de}$ 
the completely positive map 
$\cE_{\de}:\cL(\cH)\otimes\cL(\cK)\to\cL(\cH)$
defined by
$\Tr[\cE_{\de}[T]\rh] ={\langle}T,\rho\otimes \de{\rangle}$ for all normal state $\rh$ and $T\in\cL(\cH)\otimes \cL(\cK)$. From Eq.(2.1), for any 
$f, g\in C(\oR)$, we have
\beqa\label{eq:(5.2)}
\lefteqn{
\int_{\oR^2} f(x)g(y) \Pr\{[Q\in dx,Y\in dy\|U(\rho\otimes\de)U^*\}}
\quad\nn\\
&=&{\langle}g(Y)\otimes f(Q) , U(\rh\otimes\de)U^*{\rangle}\nn\\ 
&=&\Tr[\cE_{\de}[ U^*(g(Y)\otimes f (Q))U]\rh]
\eeqa
for all normal state $\rh$.

\bLemma
Let $X$ be an observable of the system I. Then for any $f\in CB(R^2)$, 
$$
\cE_\de[f(X\otimes 1,1\otimes Q)] = f (X, 0).
$$
\eLemma
\bProof
Let $\ps\in \cH$. For any $\al\in\cK$, we have 
\beqas
\lefteqn{
{\langle}\al| \cE_{\ketbra{\ps}}[f (X\otimes 1,1\otimes Q)]| \al{\rangle}}
\quad\\ 
&=&{\langle}\ps\otimes \al|f(X\otimes 1, 1\otimes Q)|\ps\otimes\al{\rangle}\\
&=&\int_{\R}\int_{\R} f(x,q) {\langle}\ps|X(dx)|\ps{\rangle}{\langle}\al|Q(dq)|\al{\rangle}\\
&=& {\langle}\al|F(Q)|\al{\rangle}, 
\eeqas
where $F(q) = \int_{R} f (x,q){\langle}\ps|X(dx)|\ps{\rangle}$. 
Thus $\cE_{\ketbra{\ps}}[f(X\otimes 1, 1\otimes Q] = F(Q)$. 
It is easy to see that $F\in CB(\R)$ and hence 
${\langle}F(Q),\de{\rangle} = F(0)$ by (D1). We see that 
\beqas
{\langle}F(Q),\de{\rangle} 
&=& 
{\langle}\cE_{\ketbra{\ps}}[f(X\otimes 1,1\otimes Q)],\de{\rangle}\\
&=&
{\langle}\ps|\cE_{\ketbra{\de}}[f(X\otimes 1,1\otimes Q)]|\ps{\rangle},
\eeqas
and
\beqas
 F(0) = {\langle}\ps|f(X,0)|\ps{\rangle}.
 \eeqas 
It follows that $\cE_{\de}[f(X\otimes 1,1\otimes Q)]  = f(X,0)$.
\eProof

\bTheorem
For any $f\in CB(\R)$, we have
$$
\cE_\de[U^*(1\otimes f(Q))U] = f(X).
$$
\eTheorem
\bProof
From Theorem 4.1, 
$U^*(1\otimes f (Q))U= f (X\otimes 1 + 1\otimes Q)$ 
and hence the assertion follows from applying Lemma 5.1 to 
$g\in CB(R^2)$ such 
that $g(x,y) = f(x + y)$.
\eProof

\bTheorem
For any $Y\in \cL(\cH)$, we have
\beq
\cE_{\de}[U^*(Y\otimes 1)U] = \cE^{X}_{\et}(Y).
\eeq
\end{Theorem} 

\begin{Proof} 
Let $\ps \in  \cH$.  For any $\al \in  \cK$, we have
\beqas
\lefteqn{{\langle}\al|\cE_{|\ps{\rangle}{\langle}\ps|}[U^*(Y\otimes 1)U]|\al{\rangle}} \\
&=& {\langle}\ps\otimes \al|U^*(Y\otimes 1)U|\ps\otimes \al{\rangle} \\
&=& \int _{\R} {\langle}\al(p)\ps|e^{ipX}Y\,
                          e^{-ipX}|\al(p)\ps{\rangle}\,dp \\
&=& \int _{\R} {\langle}\ps|e^{ipX}Y\,
                    e^{-ipX}|\ps{\rangle}\,
{\langle}\al|E^{P}(dp)|\al{\rangle} \\
    &=& {\langle}\al|F(P)|\al{\rangle},
\eeqas
	where 
$
F(p) = {\langle}\ps|e^{ipX}Y\,
            e^{-ipX}|\ps{\rangle}.
$
Consequently,
$
\cE_{|\ps{\rangle}{\langle}\ps|}[U^*(Y\otimes 1)U] = F(P).
$
Since $F\in CB(\R)$, we have from (D2), ${\langle}F(P),\de{\rangle} = \et(F)$.
We see that
\beqas
{\langle}F(P),\de{\rangle} &=&{\langle}\cE_{|\ps{\rangle}{\langle}\ps|}[U^*(Y\otimes 1)U],\de{\rangle}\\
 &=&{\langle}\psi|\cE_{\de}[U^*(Y\otimes I)U]|\psi{\rangle},
 \eeqas
 and 
 \beqas
 \et(F)&=&  \et_{p} {\langle}\ps|e^{ipX}Y\,
                  e^{-ipX}|\ps{\rangle} \\
           &=& {\langle}\ps|\cE^{X}_\et(Y)|\ps{\rangle}.
\eeqas
Thus,  $\cE_{\de}[U^*(Y\otimes 1)U] = \cE^{X}_\et(Y)$.
\end{Proof}

\bTheorem
Let $Y\in \cL(\cH)$  and $f \in  CB(\R)$.  Then we have
\beq
\cE_{\de}[U^*(Y\otimes f(Q))U] = f(X)\cE^{X}_\et[Y].
\eeq
\end{Theorem}
\begin{Proof}
By the Stinespring theorem [12], there is a Hilbert space
$\cW$, an isometry $V: \cH\otimes \cK \to  \cW$ and a $^*$-representation
$\pi: \cL(\cH)\otimes \cL(\cK) \to  \cL(\cW)$ such that $\cE_{\de}[U^*AU] =
V^*\pi (A)V$ for all $A \in  \cL(\cH)\otimes \cL(\cK)$.  
By Theorem 5.2,
$V^*\pi (1\otimes f(Q))V = f(X)$.  Thus by easy computations,
$$
(\pi (1\otimes f(Q))V - Vf(X))^* (\pi (1\otimes f(Q))V - Vf(X)) = 0.
$$
It follows that $\pi (1\otimes f(Q))V = Vf(X)$, and hence from
Theorem 5.3, we have
\beqas
\cE_{\de}[U^*(Y\otimes f(Q))U] &=& V^*\pi (Y\otimes f(Q))V 
= V^*\pi (Y\otimes 1)Vf(X) \\
&=& \cE^{X}_\et[Y]f(X) 
= f(X)\cE^{X}_\et[Y].
\eeqas
\end{Proof}

     Now we can prove that the canonical measuring process of $X$  with
preparation $\de$, where $\de$ is an $\et$-Dirac state,
satisfies the Srinivas collapse postulate for the given
invariant mean $\et$.

\bTheorem
For any bounded observable $Y\in\cL(\cH)$  and $B\in\cB(\R)$, we have
\beqas
\int_{\R} y \Pr\{X\in B; Y\in dy\|\rh\}
=\Tr[X(B) \cE^{X}_\et[Y]\rh]
\eeqas
for all normal state $\rho$.
\end{Theorem}
\begin{Proof}
Denote by $C_0(\R)$ the space of continuous functions on $\R$
vanishing at infinity. Let $Y$ be a bounded observable and $\rh$  a
normal state. From Eqs.~(5.1) and (5.2) and from Theorem 5.4, for any
$f, g \in C_{0}(\R)$ we have
$$
\int_{\oR^2} f(x)g(y) \Pr\{X\in dx;Y\in dy\|\rho\} 
= \Tr [f ( X ) \cE^X_{\eta} [g (Y ) ] \rh] .
$$
By the bounded convergence theorem and the normality of the state $\rh$,
the set of all Borel functions $f$ satisfying the above equality
is closed under bounded pointwise convergence and contains $C_{0}(\R)$.
Thus the equality holds for all bounded Borel functions $f$. Since $Y$
is bounded, there is a function $h\in C_{0}(\R)$ such that $h(y)= y$ on
the spectrum of Y.  Let $f=\chi_ B$ and $g=h$. We have $f(X)=X(B)$
and $f(Y)= Y$ so that we obtain the desired equality. 
\eProof

\section*{References}
\renewcommand{\labelenumi}{[\arabic{enumi}]}
\begin{enumerate}
\itemsep=0in
\item E.B. Davies: Quantum Theory of Open Systems, 
Academic Press, London (1976).

\item F.P. Greenleaf: Invariant Means on Topological Groups, 
Van Nostrand, New York (1969). 

\item A.S. Holevo: Probabilistic and Statistical Aspects of Quantum Theory, 
North-Holland, Amsterdam (1982). 

\item G. L\"{u}ders: \"{U}ber die Zustands\"{a}nderung durch den Messprozess, 
Ann.\ Physik 8 (1951), 322--328. 

\item J. von Neumann: Mathematical Foundations of Quantum Mechanics, 
Princeton U.P., Princeton (1955). 

\sloppy
\item M. Ozawa: Conditional expectation and repeated measurements of 
continuous quantum observables, 
Lecture Notes in Math.\ 1021 (1983), 518--525. 

\item M. Ozawa: Quantum measuring processes of continuous observables, 
J. Math.\ Phys.\ 25 (1984), 79--87. 

\item M. Ozawa: Conditional probability and a posteriori states in quantum mechanics, 
Publ.\ Res.\ Inst.\ Math.\ Sci., Kyoto Univ. 21 (1985), 279--295. 

\item M. Ozawa: Concepts of conditional expectations in quantum theory, 
J. Math.\ Phys.\ 26 (1985),1948--1955. 

\item M. Ozawa: On information gain by quantum measurements of continuous observables, 
J. Math.\ Phys.\ 27 (1986), 759--763. 

\item M.D. Srinivas: Collapse postulate for observables with continuous spectra, 
Commun.\ Math.\ Phys. 71 (1980), 131--158. 

\item W.F. Stinespring: Positive functions on C*-algebras, 
Proc.\ Amer.\ Math. Soc. 6 (1955), 211--216. 
\end{enumerate}
\end{document}